\documentclass[a4paper,12pt]{article}
\usepackage[pctex32]{graphics}
\usepackage{amssymb,amsmath}

\begin{document}
\topmargin 0pt \oddsidemargin 0mm
\newcommand{\be}{\begin{equation}}
\newcommand{\ee}{\end{equation}}
\newcommand{\ba}{\begin{eqnarray}}
\newcommand{\ea}{\end{eqnarray}}
\newcommand{\fr}{\frac}
\renewcommand{\thefootnote}{\fnsymbol{footnote}}
\def \nn {\nonumber}
\def\a{\alpha}      \def\da{{\dot\alpha}}
\def\b{\beta}       \def\db{{\dot\beta}}
\def\c{\gamma}  \def\C{\Gamma}  \def\cdt{\dot\gamma}
\def\d{\delta}  \def\D{\Delta}  \def\ddt{\dot\delta}
\def\e{\epsilon}        \def\vare{\varepsilon}
\def\f{\phi}    \def\F{\Phi}    \def\vvf{\f}
\def\h{\eta}
\def\k{\kappa}
\def\l{\lambda} \def\L{\Lambda}
\def\m{\mu} \def\n{\nu}
\def\o{\omega}
\def\P{\Pi}
\def\r{\rho}
\def\s{\sigma}  \def\S{\Sigma}
\def\t{\tau}
\def\th{\theta} \def\Th{\Theta} \def\vth{\vartheta}
\def\X{\Xeta}
\def\z{\zeta}
\def\w{\wedge}
\def\u{\underline}
\def\hs{\hspace}
\begin{titlepage}

\vspace{5mm}
\begin{center}
{\Large \bf Logarithmic quasinormal modes of a  spin-3 field around
the BTZ black hole}

\vskip .6cm

\centerline{\large
 Yong-Wan Kim $^{1,a}$,  Yun Soo Myung$^{1,b}$,
and Young-Jai Park$^{2,3,c}$}

\vskip .6cm

{$^{1}$Institute of Basic Science and School of Computer Aided
Science, \\Inje University, Gimhae 621-749, Korea \\}

{$^{2}$Department of Physics and Center for Quantum Spacetime, \\
Sogang University, Seoul 121-742, Korea}\\

{$^{3}$Department of Global Service Management,\\
Sogang University, Seoul 121-742, Korea}

\end{center}

\begin{center}

\underline{Abstract}
\end{center}
Using the operator approach, we obtain  logarithmic quasinormal
modes and frequencies of a traceless spin-3 field  around the BTZ
black hole at the critical point of the spin-3 topologically
massive gravity. The logarithmic quasinormal frequencies are also
confirmed by considering logarithmic conformal field theory.

\vskip .6cm



\vskip 0.8cm

\vspace{15pt} \baselineskip=18pt

\noindent $^a$ywkim65@gmail.com\\
\noindent $^b$ysmyung@inje.ac.kr \\
\noindent $^c$yjpark@sogang.ac.kr

\thispagestyle{empty}
\end{titlepage}

\newpage
\section{Introduction}

Recently,  higher-spin theories on the (2+1)-dimensional anti-de
Sitter (AdS$_3$) spacetimes have been the subject of active
interest because they admit a truncation  to an arbitrary maximal
spin $N$~\cite{henneaux,theisen}. Especially, the  prototype of
spin-3 model is a totally symmetric third-rank tensor of spin-3
field coupled to topologically massive gravity (TMG). Chen {\it et
al.,}~\cite{CLW} have developed the methods obtaining quasinormal
modes of arbitrary spin theories, and  discussed the traceless
spin-3 fluctuations around the AdS$_3$ spacetimes. They found that
there exists a single massive propagating mode, besides
left-moving and right-moving massless modes (gauge artifacts).
On the other hand, Bagchi {\it et al.,}~\cite{BLSS} have
independently studied the spin-3 TMG. They showed that the trace
modes carry energy opposite in sign to the traceless modes, and
pointed out the instability in the bulk of the logarithmic partner
of the traceless modes. These are considered through extended
analysis of spin-2 field in the cosmological TMG~\cite{LSS}.

Very recently, Datta and David~\cite{DD} have introduced  massive
wave equations of arbitrary integer spin fields including spin-3
fields in the Ba\~nados-Teitelboim-Zanelli (BTZ) black hole
background. Then, they have obtained their quasinormal modes that
are consistent with the location of the poles of the corresponding
two-point function in the dual conformal field theory. This could
be predicted by the AdS$_3$/CFT$_2$ correspondence.
By the way, they have solved the second-order perturbed equation of
$[\bar{\Box}-m^2+4/\ell^2] \Phi_{\rho\mu\nu}=0$ for spin-3 field
with the ingoing modes at horizon and Dirichlet boundary condition
at infinity. However, in this case, there exists a sign ambiguity
of mass $\pm m$. Thus, in order to avoid this ambiguity, one has
to  solve the first-order equation  of
$\epsilon_\rho^{~\alpha\beta}\bar{\nabla}_\alpha\Phi_{\beta\mu\nu}+m\Phi_{\rho\mu\nu}=0$
itself under  the transverse and traceless (TT) gauge condition.

It was known that the operator approach (method)~\cite{SS,BMS} is
very useful to derive the quasinormal modes of spin-2 field of
graviton in the BTZ black hole background in the framework of the
cosmological TMG. This method has been applied to the new massive
gravity to derive their quasinormal modes of the BTZ black
hole~\cite{MKMP}. Very recently, we  have obtained quasinormal
modes of the BTZ black hole in spin-3 TMG by  using the operator
method~\cite{MKP}. This method shows clearly how to derive
quasinormal modes without any sign ambiguity in mass.

On the other hand, the presence of the logarithmic modes at the
critical point of the TMG was pointed
out~\cite{GJ1,Giribet:2008bw,Myunglcft}. In particular, Grumiller
and Johansson~\cite{GJ1} have shown that these modes grow linearly
in time and the radial coordinate of the AdS$_3$ spacetimes, which
cause issues on the stability and the chiral nature of the theory.
After their work, a derivation of the logarithmic quasinormal
modes of spin-2 was performed  for the BTZ back hole
~\cite{Sachs}. It seems that the operator approach is the only
known method to derive the logarithmic quasinormal modes of spin-3
field because solving  the second-order equation at the critical
point cannot provide appropriate logarithmic quasinormal modes, in
compared with the non-critical case.

In this work, we wish to derive logarithmic quasinormal modes and
frequencies of a traceless spin-3 field around the BTZ black hole
at the critical point of spin-3 topologically massive gravity
theory. We will observe  how they differ from the logarithmic
quasinormal modes of a spin-2 field. Also, we explore Log-boundary
conditions for left-and right-logarithmic modes. Finally, these
quasinormal frequencies will be also confirmed by considering a logarithmic
conformal field theory (LCFT).


\section{Perturbation analysis for spin-3 field}


\subsection{Action of spin-3 TMG}
The action  for  spin-3 coupled to TMG is given by
 \ba\label{spin3-action}
 S&=&\frac1{8\pi G}\int\Big[ e^a\w
 d\omega_a+\frac12 \epsilon_{abc}e^a\w
 \omega^b\w\omega^c+\frac1{6l^2}\epsilon_{abc}e^a\w e^b \w e^c
 -2\sigma e^{ab}\w d\omega_{ab}\nn\\
 &&~~~~~~~~~~~~-2\sigma\epsilon_{abc}e^a\w\omega^{bd}\w\omega^c_{\,\,d}-2\sigma
 e^{ab}\w\epsilon_{(a|cd}\omega^c\w \omega^{\,\,\,\,d}_{|b)}
 -\frac{2\sigma}{l^2}\epsilon_{abc}e^a\w e^{bd}\w e^c_{\,\,d}\Big]
 \nn \\
 &&- \frac1{16\pi G\mu}\int \Big[ \omega^a\w d\omega_a+\frac13
 \epsilon_{abc}\omega^a\w\omega^b\w\omega^c -2\sigma\omega^{ab}\w
 d\omega_{ab}-4\sigma\epsilon_{abc}\omega^a\w\omega^{bd}\w\omega^c_{\,\,d}
 \nn\\
 &&~~~~~~~~~~~~~~~~~+\beta^a\w T_a-2\sigma \beta^{ab}\w T_{ab}\Big],\label{action}
 \ea
where $\sigma<0$ is a free parameter from SL(3,R) gauge group. Here
two Lagrange multipliers
 \be
 \b^{a}=\tilde\b^{a}+\frac{e^{a}}{l^2},~~~\b^{ab}=\tilde\b^{ab}+\frac{e^{ab}}{l^2}
 \ee
are introduced to impose the torsion free conditions~\cite{theisen}
 \ba
 T^a &\equiv& de^a+\epsilon^{abc}\omega_b\w
         e_c-4\sigma\epsilon^{abc}e_{bd}\w\omega^{~d}_c=0,\nonumber\\
 T^{ab} &\equiv& de^{ab}+\epsilon^{cd(a|}\omega_c\w e^{~|b)}_d
            +\epsilon^{cd(a|}e_c\w\omega^{~|b)}_d=0.
 \ea
The former in Eq. (\ref{spin3-action}) denotes the action for the
spin-3 AdS$_3$ gravity~\cite{theisen}, while the latter represents
the spin-3 generalization of gravitational Chern-Simons term with
a coupling constant $1/\mu$. The equations of motion obtained by
varying this action are given by the torsion free conditions \ba
T^a=0, \hs{3ex} T^{ab}=0,\ea and four equations \ba &&
R_a-\frac1{2\mu }(d\beta_a+\epsilon_{abc}\beta^b\w
\omega^c-2\sigma\epsilon_{(c|da}\beta^{bc}\w \omega^d_{\,\,|b)})=0,\\
&& R_a+\frac12\epsilon_{abc}\left[\beta^b\w e^c-\frac{e^b\w
e^c}{l^2}+4\sigma\left(\frac{e^{bd}\w e^c_{\,\,d}}{l^2}-e^{bd}\w
\beta^c_{\,\,d}\right)\right]=0,\\
&& R_{ab}-\frac1{2\mu}\left(d\beta_{ab}+\epsilon_{cd(a|}\beta^c\w
\omega^d_{\,\,|b)}+\epsilon_{cd(a|}\omega^c\w
\beta^d_{\,\,|b)}\right)=0,\\
&& R_{ab}+\frac12\left(\epsilon_{cd(a|}\beta^c\w
e^d_{\,\,|b)}+\epsilon_{cd(a|}e^c\w
\beta^d_{\,\,|b)}\right)-\frac1{l^2}\epsilon_{cd(a|}e^c\w
e^d_{\,\,|b)}=0
\ea
with
\ba
R_a&=&d\omega_a+\frac12\epsilon_{abc}(\omega^b\w
\omega^c+\frac{e^b\w e^c}{l^2} )-2\sigma\epsilon_{abc}(\omega_{bd}\w
\omega_c^{\,\,d}+\frac{e_{bd}\w e_c^{\,\,d}}{l^2}),\\
R_{ab}&=&d\omega_{ab}+\epsilon_{cd(a|}\omega^c\w
\omega^d_{\,\,|b)}+\frac{1}{l^2}\epsilon_{cd(a|}e^c\w e^d_{\,\,|b)}.
\ea At this stage, we note that these  differ from the pure gravity
coupled to spin-3 field theory  by $\b^a$ and $\b^{ab}$ terms.
However, for  \be \b^a=\frac{e^a}{l^2}, \hs{3ex}
\b^{ab}=\frac{e^{ab}}{l^2}, \ee the extra terms disappear due to the
torsion free conditions, leading to the pure gravity coupled to
spin-3 field theory~\cite{theisen}. This implies that the
nonrotating BTZ black hole solution to pure gravity coupled to
spin-3 field theory~\cite{BCPP}
 \be
 \bar{e}^a=e^a_{\rm BTZ}
 \ee
with
 \be
 e^0_{\rm BTZ}=\left(-M+\frac{r^2}{l^2}\right)dt,
 ~~e^1_{\rm BTZ}=\left(-M+\frac{r^2}{l^2}\right)^{-1}dr,
 ~~e^2_{\rm BTZ}=rd\phi
 \ee
is also the solution to the above equations of motion. Here, the
spin connection $\bar{\omega}^a=\bar{\omega}^a_{\rm BTZ}$ takes its
components as \be \bar{\omega}^0_{\rm
BTZ}=\frac{1}{r}\left(-M+\frac{r^2}{l^2}\right)
\bar{e}^2,~~\bar{\omega}_{\rm BTZ}^1=0,~~\bar{\omega}^2_{\rm
BTZ}=\frac{\partial}{\partial
r}\left(-M+\frac{r^2}{l^2}\right)\bar{e}^0. \ee In addition, one has
\be \bar{\beta}^a=\frac{e^a_{\rm BTZ}}{l^2},~~
\bar{e}^{ab}=e^{ab}_{\rm BTZ}=0,~~
\bar{\omega}^{ab}=\omega^{ab}_{\rm
BTZ}=0,~~\bar{\beta}^{ab}=\beta^{ab}_{\rm BTZ}=\frac{e^{ab}_{\rm
BTZ}}{l^2}=0 \ee for the BTZ black hole.

\subsection{Perturbation for spin-3 field}
Now we consider the perturbations around the BTZ black hole
background with background fields
$\bar{e}^a,~\bar{\omega}^a,~\bar{\beta}^a,~\bar{e}^{ab},~\bar{\omega}^{ab},$
and $\bar{\beta}^{ab}$.
 For simplicity,  we  denote the perturbed
fields as $e^a, \cdots$ without the bar notation ( $\bar{}$ ). The
six perturbed equations take the forms \ba
de^a+\epsilon^{abc}\bar\omega_b\w
e_c+\epsilon^{abc}\omega_b\w \bar{e}_c=0,\\
d\omega_a+\epsilon_{abc}(\bar\omega^b\w \omega^c+\frac{\bar{e}^b\w
e^c}{l^2})-\frac1{2\mu }(d\beta_a+\epsilon_{abc}\bar\beta^b\w
\omega^c+\epsilon_{abc}\beta^b\w \bar\omega^c)=0,\\
d\omega_a+\epsilon_{abc}(\bar\omega^b\w \omega^c+\frac{\bar{e}^b\w
e^c}{l^2})+\frac12\epsilon_{abc}\left[\bar\beta^b\w e^c+\beta^b\w
\bar{e}^c-\frac{2}{l^2}\bar{e}^b\w e^c\right]=0,\\
de^{ab}+\epsilon^{cd(a|}\bar{\omega}_c\w
e_d^{\,\,|b)}+\epsilon^{cd(a|} \bar{e}_c\w
\omega_d^{\,\,|b)}=0,\label{torsion2}\\
R_{ab}-\frac1{2\mu}\left(d\beta_{ab}+\epsilon_{cd(a|}\bar\beta^c\w
\omega^d_{\,\,|b)}+\epsilon_{cd(a|}\bar\omega^c\w
\beta^d_{\,\,|b)}\right)=0,\label{eom2}\\
R_{ab}+\frac12\left(\epsilon_{cd(a|}\bar\beta^c\w
e^d_{\,\,|b)}+\epsilon_{cd(a|}\bar{e}^c\w
\beta^d_{\,\,|b)}\right)-\frac1{l^2}\epsilon_{cd(a|}\bar{e}^c\w
e^d_{\,\,|b)}=0 \label{eom1} \ea with the perturbed Ricci tensor
\ba R_{ab}=d\omega_{ab}+\epsilon_{cd(a|}\bar\omega^c\w
\omega^d_{\,\,|b)}+\frac{1}{l^2}\epsilon_{cd(a|}\bar{e}^c\w
e^d_{\,\,|b)}. \ea

Hereafter, we will express the perturbed fields in terms of the
frame fields $h_{\mu\nu}$ and $\Phi_{\mu\nu\lambda}$ as \be
h_{\mu\nu}= e_{\mu a}\bar{e}^a_\nu,~~~\Phi_{\mu\nu\lambda}=e_{\mu
ab}\bar{e}^a_\nu \bar{e}^b_\lambda, \ee where the Latin indices of
$a$ and $b$ are replaced by the Greek indices of $\nu$ and
$\lambda$. The Greek indices are raised (or lowered) by the BTZ
metric of $\bar{g}^{\rm BTZ}_{\mu\nu}=\bar{e}^a_\mu \bar{e}^b_\nu
\eta_{ab}$ where \be ds^2_{\rm BTZ}=\bar{g}^{\rm
BTZ}_{\mu\nu}dx^\mu
dx^\nu=-\left(-M+\frac{r^2}{l^2}\right)dt^2+\frac{dr^2}{\left(-M+\frac{r^2}{l^2}\right)}+r^2d\phi^2.
\ee

The perturbed equation of spin-2 graviton takes the form~\cite{LSS}
\be
 \Big(\bar{\Box}+\frac{2}{l^2}\Big)h^{\rho}_{~\sigma}+\frac{1}{\mu}\epsilon^{\rho\mu\nu}\bar{\nabla}_{\mu}
 \Big(\bar{\Box}+\frac{2}{l^2}\Big)h_{\nu\sigma}=0,\label{spin2eq}
 \ee
which  is  decoupled completely from the spin-3 perturbed equation
as~\cite{CLW}
 \be \bar{\Box}
 \Phi^{\rho\alpha\beta}+\frac1{2\mu}\epsilon^{\rho\mu\nu}\bar{\nabla}_\mu
 \bar{\Box}\Phi_\nu^{\,\,\,\alpha\beta}=0.\label{spin3eq}
 \ee

In this work, we consider the  BTZ black hole with
 the mass $M=1$ and the AdS$_3$ curvature radius $l=1$ in global coordinates as
\begin{equation}\label{metric}
 ds^2_{\rm BTZ} = \bar{g}_{\mu\nu}dx^\mu dx^\nu
      = -\sinh^2\!\!\rho\, d\tau^2+\cosh^2\!\!\rho\, d\phi^2+d\rho^2,
\end{equation}
where the event horizon is located at $\rho=0$, while the infinity
is at $\rho=\infty$. Introducing  the light-cone coordinates
$u/v=\tau\pm \phi$, the metric tensor $\bar{g}_{\mu\nu}$ takes the
form
\begin{equation} \bar{g}_{\mu\nu}=
\left(
  \begin{array}{ccc}
    \frac{1}{4} & -\frac{1}{4}\cosh\!2\rho & 0 \\
      -\frac{1}{4}\cosh\!2\rho & \frac{1}{4}  & 0 \\
    0 & 0 & 1 \\
  \end{array}
\right). \label{newm}
\end{equation}
Then the metric tensor (\ref{newm}) admits the Killing vector
fields $L_k$ $(k=0,-1,1)$ for the local SL(2,R)$\times$SL(2,R)
algebra as
\begin{equation}
L_0=-\partial_u,~~L_{-1/1}=e^{\mp
u}\Bigg[-\frac{\cosh\!2\rho}{\sinh\!2\rho}\partial_u-\frac{1}{\sinh\!2\rho}\partial_v
\mp \frac{1}{2} \partial_\rho\Bigg],
\end{equation}
and $\bar{L}_0$ and $\bar{L}_{-1/1}$ are obtained by substituting
$u\leftrightarrow v$. Locally, they form a basis of the SL(2,R)
Lie algebra  as
\begin{equation}
[L_0,L_{\pm 1}]=\mp L_{\pm 1},~~[L_1,L_{-1}]=2L_0.
\end{equation}
Here, we note that the totally symmetric spin-3 field
$\Phi^{\rho\mu\nu}$ satisfying the TT gauge condition
 \be\label{ttg}
 \bar{\nabla}^\mu\Phi_{\mu\nu\rho}=0,~~~\Phi_\mu^{~\mu\nu}=0
 \ee
has only one degree of freedom corresponding to a single
massive mode propagating in the BTZ black hole
spacetimes~\cite{MKP}.

On the other hand, the third-order equation~(\ref{spin3eq}) can also be expressed as
 \be\label{3rdeq2}
 ({\cal D}^M{\cal D}^L{\cal D}^R \Phi)^{\rho\mu\nu}=0
 \ee
in terms of mutually commuting operators
 \ba\label{mco}
 ({\cal D}^{L/R})^{\rho\nu}=\delta^{\rho\nu}\pm \frac{1}{2}\epsilon^{\rho\mu\nu}\bar{\nabla}_\mu,
 ~~({\cal D}^M)^{\rho\nu}=\delta^{\rho\nu}+\frac{1}{2\mu}\epsilon^{\rho\mu\nu}\bar{\nabla}_\mu.
 \ea
We note that Eq. (\ref{3rdeq2}) is reduced to Eq. (\ref{spin3eq}) when using  the BTZ background
 \ba
 \bar{R}_{\rho\sigma\mu\nu}=-(\bar{g}_{\rho\mu}\bar{g}_{\sigma\nu}-\bar{g}_{\rho\nu}\bar{g}_{\sigma\mu}),
 ~~~\bar{R}_{\mu\nu=}=-2\bar{g}_{\mu\nu},
  \ea
together with the TT gauge condition and the relation of
$[\bar{\nabla}_\mu,\bar{\nabla}_\nu]
\Phi^{\mu\alpha\beta}=-4\Phi_\nu^{~\alpha\beta}$. Therefore, the
third-order equation~(\ref{spin3eq}) can be decomposed into three
first-order differential equations:
 \be\label{three1st}
 ({\cal D}^M \Phi)^{\rho\mu\nu}=0,~~({\cal D}^L
 \Phi)^{\rho\mu\nu}=0,~~({\cal D}^R \Phi)^{\rho\mu\nu}=0,
 \ee
for a massive, a left-moving, and a right-moving degree of freedom, respectively.
Importantly, three first-order differential
equations~(\ref{three1st}) can be simply rewritten in terms of a
single massive first-order differential equation as
 \be\label{meof}
 \epsilon_\rho^{~\alpha\beta}\bar{\nabla}_\alpha\Phi_{\beta\mu\nu}+m\Phi_{\rho\mu\nu}=0
 \ee
with $m=2\mu,~2$, and $-2$.

It seems appropriate to comment that it could also be
expressed in terms of a second-order differential
equation~\cite{DD} as
 \be \label{meos}
 \Big[\bar{\Box}^2-m^2+4\Big]\Phi_{\rho\mu\nu}=0.
 \ee
However, there exists a sign ambiguity $\pm m$ in this equation.
Therefore, in order to avoid this ambiguity, we will  directly solve the first-order equation
(\ref{meof}) with the TT gauge condition.
Note that at the chiral (critical) point of $\mu=1$, the operators ${\cal
D}^M$ and ${\cal D}^L$ degenerate.

Having the structure in mind, let us find quasinormal modes for the
spin-3 field in the BTZ background by solving (\ref{meof}) together
with the TT gauge condition. In order to implement the operator
method~\cite{SS,MKMP}, one  has to choose either the anti-chiral
highest weight condition of $L_1\Phi_{\rho\mu\nu}=0$ or the chiral
highest weight condition of $\bar{L}_1\Phi_{\rho\mu\nu}=0$, but not
both simultaneously. Actually,   we note that   for a generic
symmetric tensor $\Phi_{\rho\mu\nu}$, the transversality condition
of $\bar{\nabla}^\mu\Phi_{\mu\nu\rho}=0$ is not equivalent to
choosing the chiral (anti-chiral) highest weight condition. However,
selecting  proper components of $\Phi_{\rho\mu\nu}$, two are
equivalent to each other.


\section{Left-logarithmic quasinormal modes}

We observe that similar to the perturbed equation (\ref{spin2eq})
for the spin-2 graviton, the spin-3 fluctuation also satisfies a
third-order differential equation (\ref{spin3eq}). Since the
logarithmic quasinormal modes of the graviton were computed at the
critical point in~\cite{Sachs}, we wish to calculate logarithmic
quasinormal modes of the spin-3 field  by the operator method
in this section.


\subsection{Logarithmic quasinormal modes}
By solving the first-order equation (\ref{meof}), we obtain the
left-moving solution of a anti-chiral highest weight field
 \be\label{leftsol01}
 \Phi^L_{\rho\mu\nu}(u,v,\rho)=e^{ik(t-\phi)-2h_L(m)t}(\sinh\rho)^{-2h_L(m)}(\tanh\rho)^{ik}F^L_{\rho\mu\nu}(\rho),
 \ee
where $h_L(m)=(m-2)/2$ and $F^L_{\rho\mu\nu}(\rho)$ is given
by~\cite{MKP}
 \ba\label{leftm1}
 F^L_{u\mu\nu}(\rho)&=&\left(
  \begin{array}{ccc}
    0 & 0 & 0\\
    0 & 0 & 0 \\
    0 & 0 & 0 \\
  \end{array}
 \right)_{\mu\nu},  \nonumber\\
  \label{leftm2}
 F^L_{v\mu\nu}(\rho)&=&\left(
  \begin{array}{ccc}
    0 & 0 & 0\\
    0 & 1 & \frac{2}{\sinh2\rho} \\
    0 & \frac{2}{\sinh2\rho} & \frac{4}{\sinh^22\rho} \\
  \end{array}
 \right)_{\mu\nu}, \nonumber\\
  \label{leftm3}
 F^L_{\rho\mu\nu}(\rho)&=&\left(
  \begin{array}{ccc}
    0 & 0 & 0\\
    0 & \frac{2}{\sinh2\rho} & \frac{4}{\sinh^22\rho}\\
    0 & \frac{4}{\sinh^22\rho} & \frac{8}{\sinh^32\rho} \\
  \end{array}
 \right)_{\mu\nu}.
 \ea
Note here that $F^L_{v\mu\nu}(\rho)$ takes the same form as the
spin-2 graviton  which is Eq. (17) in Ref.~\cite{BMS}, while
$F^L_{u\mu\nu}(\rho)$ is null and $F^L_{\rho\mu\nu}(\rho)$ is more
damped than $F^L_{v\mu\nu}(\rho)$ for large $\rho$.

However, the basis of solutions (\ref{leftsol01}) becomes inadequate at the chiral point $\mu = 1$,
since the $L$ and $M$ branches coincide at this point from Eq. (\ref{mco}).
One could remedy this problem by constructing a new mode solution
satisfying~\cite{GJ1}
 \be
 {\cal D}^L \Phi^{L,new}_{\rho\mu\nu}=-\Phi^L_{\rho\mu\nu}\neq 0,
 \ee
so that
 \be
  {\cal D}^L {\cal D}^L \Phi^{L,new}_{\rho\mu\nu}=- {\cal D}^L  \Phi^L_{\rho\mu\nu}=0,
 \ee
while satisfying with the anti-chiral highest weight condition as
well as the TT gauge condition.

Then, at the critical point of $\mu=1$ ($m=2$), a new logarithmic solution is given by
 \ba
 \Phi^{L,new}_{\rho\mu\nu}&=&\left.\partial_m \Phi^{L,m}_{\rho\mu\nu}\right|_{m\rightarrow 2}\nonumber\\
                        &=&   y(t,\rho) \left.\Phi^L_{\rho\mu\nu}(u,v,\rho)\right|_{m\rightarrow 2},
 \ea
where
 \be\label{yftn}
 y(t,\rho)=-t-\ln\sinh\rho.
 \ee
Here, $\left.\Phi^L_{\rho\mu\nu}(u,v,\rho)\right|_{m\rightarrow
2}$ is given by Eq. (\ref{leftsol01}) as
 \be\label{nsol}
 \left.\Phi^L_{\rho\mu\nu}(u,v,\rho)\right|_{m\rightarrow 2}=e^{ik(t-\phi)}(\tanh\rho)^{ik}F^L_{\rho\mu\nu}(\rho).
 \ee

Next, according to Sachs's proposal~\cite{Sachs},
the logarithmic quasinormal modes can be constructed by using the following operation
 \be
 \Phi^{L(n),new}_{\rho\mu\nu}(u,v,\rho)=
 \Big(\bar{L}_{-1}L_{-1}\Big)^n
 \Phi^{L,new}_{\rho\mu\nu}(u,v,\rho),
 \ee
which means that we should compute their descendants of
$\Phi^{L,new}_{\rho\mu\nu}(u,v,\rho)$ by using the operator method.

The first descendants of $\Phi^{L,new}_{\rho\mu\nu}(u,v,\rho)$
represented by
 \be\label{leftsol1}
 \Phi^{L(1),new}_{\rho\mu\nu}(u,v,\rho)= \Big(\bar{L}_{-1}L_{-1}\Big)
 \Phi^{L,new}_{\rho\mu\nu}(u,v,\rho)
 \ee
are explicitly given by
 \ba
 \Big(\bar{L}_{-1}L_{-1}\Big) \Phi^{L,new}_{u\mu\nu}(u,v,\rho)
 &=&\frac{e^{-2t}}{\sinh^2\!\rho}e^{ikv}(\tanh\!\rho)^{ik}
  \left(
  \begin{array}{ccc}
    0 & 0 & 0\\
    0 & f^{L(1)}_{uvv} & \frac{2f^{L(1)}_{uvv}}{\sinh2\rho} \\
    0 & \frac{2f^{L(1)}_{uvv}}{\sinh2\rho} & \frac{4f^{L(1)}_{uvv}}{\sinh^22\rho} \\
  \end{array}  \right)_{\mu\nu}, \\
 \Big(\bar{L}_{-1}L_{-1}\Big) \Phi^{L,new}_{v\mu\nu}(u,v,\rho)
 &=&\frac{e^{-2t}}{\sinh^2\!\rho}e^{ikv}(\tanh\!\rho)^{ik}
  \left(
  \begin{array}{ccc}
    0 & f^{L(1)}_{uvv} & \frac{2f^{L(1)}_{uvv}}{\sinh2\rho}\\
    f^{L(1)}_{uvv} & \frac{f^{L(1)}_{vvv}}{2} &  \frac{f^{L(1)}_{vv\rho}}{\sinh2\rho} \\
    \frac{2f^{L(1)}_{uvv}}{\sinh2\rho} &  \frac{f^{L(1)}_{vv\rho}}{\sinh2\rho} &  \frac{2f^{L(1)}_{v\rho\rho}}{\sinh^22\rho} \\
  \end{array} \right)_{\mu\nu},\\
 \Big(\bar{L}_{-1}L_{-1}\Big) \Phi^{L,new}_{\rho\mu\nu}(u,v,\rho)
 &=&\frac{e^{-2t}}{\sinh^2\!\rho}e^{ikv}(\tanh\!\rho)^{ik}
  \left(
  \begin{array}{ccc}
    0 & \frac{2f^{L(1)}_{uvv}}{\sinh2\rho} & \frac{4f^{L(1)}_{uvv}}{\sinh^22\rho}\\
    \frac{2f^{L(1)}_{uvv}}{\sinh2\rho} & \frac{f^{L(1)}_{vv\rho}}{\sinh2\rho} &  \frac{2f^{L(1)}_{v\rho\rho}}{\sinh^22\rho} \\
    \frac{4f^{L(1)}_{uvv}}{\sinh^22\rho} & \frac{2f^{L(1)}_{v\rho\rho}}{\sinh^22\rho} &  \frac{4f^{L(1)}_{\rho\rho\rho}}{\sinh^32\rho} \\
  \end{array}
 \right)_{\mu\nu},
 \ea
whose relevant components are given by
 \ba
 f^{L(1)}_{uvv} &=& 1+(3-ik)y(t,\rho), \nonumber\\
 f^{L(1)}_{vvv}&=& 5-3ik+(3-ik)\cosh2\rho+2(3-4ik-k^2)y(t,\rho), \nonumber\\
 f^{L(1)}_{vv\rho}&=& 5-3ik+(5-ik)\cosh2\rho+2(3-4ik-k^2+(3-ik)\cosh2\rho)y(t,\rho), \nonumber \\
 f^{L(1)}_{v\rho\rho}&=& 5-3ik+(7-ik)\cosh2\rho+2(3-4ik-k^2+2(3-ik)\cosh2\rho)y(t,\rho), \nonumber \\
 f^{L(1)}_{\rho\rho\rho}&=& 5-3ik+(9-ik)\cosh2\rho+2(3-4ik-k^2+3(3-ik)\cosh2\rho)y(t,\rho).
 \ea

Then, the second descendants of $\Phi^{L,new}_{\rho\mu\nu}(u,v,\rho)$ are read off
from the operation
 \be\label{leftsol2}
 \Phi^{L(2),new}_{\rho\mu\nu}(u,v,\rho)= \Big(\bar{L}_{-1}L_{-1}\Big)^2
 \Phi^{L,new}_{\rho\mu\nu}(u,v,\rho).
 \ee
We have their explicit forms
 \ba\label{2ndDL1}
 \Big(\bar{L}_{-1}L_{-1}\Big)^2 \Phi^{L,new}_{u\mu\nu}(u,v,\rho)
 &=&\frac{e^{-4t}}{\sinh^4\!\rho}e^{ikv}(\tanh\!\rho)^{ik}
  \left(
  \begin{array}{ccc}
    0 & f^{L(2)}_{uuv} & \frac{f^{L(2)}_{uu\rho}}{\sinh2\rho}\\
    f^{L(2)}_{uuv} & f^{L(2)}_{uvv} & \frac{f^{L(2)}_{uv\rho}}{\sinh2\rho} \\
    \frac{f^{L(2)}_{uu\rho}}{\sinh2\rho} & \frac{f^{L(2)}_{uv\rho}}{\sinh2\rho} & \frac{f^{L(2)}_{u\rho\rho}}{\sinh^22\rho} \\
  \end{array}  \right)_{\mu\nu}, \\
 \label{2ndDL2}
 \Big(\bar{L}_{-1}L_{-1}\Big)^2 \Phi^{L,new}_{v\mu\nu}(u,v,\rho)
 &=&\frac{e^{-4t}}{\sinh^4\!\rho}e^{ikv}(\tanh\!\rho)^{ik}
  \left(
  \begin{array}{ccc}
    f^{L(2)}_{uuv}  & f^{L(2)}_{uvv} & \frac{f^{L(2)}_{uv\rho}}{\sinh2\rho}\\
    f^{L(2)}_{uvv} & f^{L(2)}_{vvv} & \frac{f^{L(2)}_{vv\rho}}{\sinh2\rho} \\
    \frac{f^{L(2)}_{uv\rho}}{\sinh2\rho} & \frac{f^{L(2)}_{vv\rho}}{\sinh2\rho} & \frac{f^{L(2)}_{v\rho\rho}}{\sinh^22\rho} \\
  \end{array} \right)_{\mu\nu},\\
 \label{2ndDL3}
 \Big(\bar{L}_{-1}L_{-1}\Big)^2 \Phi^{L,new}_{\rho\mu\nu}(u,v,\rho)
 &=&\frac{e^{-4t}}{\sinh^4\!\rho}e^{ikv}(\tanh\!\rho)^{ik}
  \left(
  \begin{array}{ccc}
     \frac{f^{L(2)}_{uu\rho}}{\sinh2\rho} & \frac{f^{L(2)}_{uv\rho}}{\sinh2\rho} & \frac{f^{L(2)}_{u\rho\rho}}{\sinh^22\rho}\\
    \frac{f^{L(2)}_{uv\rho}}{\sinh2\rho}  & \frac{f^{L(2)}_{vv\rho}}{\sinh2\rho} & \frac{f^{L(2)}_{v\rho\rho}}{\sinh^22\rho} \\
   \frac{f^{L(2)}_{u\rho\rho}}{\sinh^22\rho} &  \frac{f^{L(2)}_{v\rho\rho}}{\sinh^22\rho}  & \frac{f^{L(2)}_{\rho\rho\rho}}{\sinh^32\rho} \\
  \end{array}
 \right)_{\mu\nu}.
 \ea
The full expressions of the matrix elements, $f^{L(2)}_{uuv}$, etc., are listed in Appendix A1.

On the the hand, the third descendants of $\Phi^{L,new}_{\rho\mu\nu}(u,v,\rho)$ are given by
 \be\label{leftsol3}
 \Phi^{L(3),new}_{\rho\mu\nu}(u,v,\rho)= \Big(\bar{L}_{-1}L_{-1}\Big)^3
 \Phi^{L,new}_{\rho\mu\nu}(u,v,\rho).
 \ee
We have
 \ba\label{3rdDL1}
 \Big(\bar{L}_{-1}L_{-1}\Big)^3 \Phi^{L,new}_{u\mu\nu}(u,v,\rho)
 &=&\frac{e^{-6t}}{\sinh^6\!\rho}e^{ikv}(\tanh\!\rho)^{ik}
  \left(
  \begin{array}{ccc}
    f^{L(3)}_{uuu}  & f^{L(3)}_{uuv} & \frac{f^{L(2)}_{uu\rho}}{\sinh2\rho}\\
    f^{L(3)}_{uuv} & f^{L(3)}_{uvv} & \frac{f^{L(3)}_{uv\rho}}{\sinh2\rho} \\
   \frac{f^{L(2)}_{uu\rho}}{\sinh2\rho} & \frac{f^{L(3)}_{uv\rho}}{\sinh2\rho} & \frac{f^{L(3)}_{u\rho\rho}}{\sinh^22\rho} \\
  \end{array}  \right)_{\mu\nu}, \\
 \label{3rdDL2}
 \Big(\bar{L}_{-1}L_{-1}\Big)^3 \Phi^{L,new}_{v\mu\nu}(u,v,\rho)
 &=&\frac{e^{-6t}}{\sinh^6\!\rho}e^{ikv}(\tanh\!\rho)^{ik}
  \left(
  \begin{array}{ccc}
    f^{L(3)}_{uuv}  & f^{L(3)}_{uvv} & \frac{f^{L(3)}_{uv\rho}}{\sinh2\rho}\\
    f^{L(3)}_{uvv} & f^{L(3)}_{vvv} & \frac{f^{L(3)}_{vv\rho}}{\sinh2\rho} \\
    \frac{f^{L(3)}_{uv\rho}}{\sinh2\rho} & \frac{f^{L(3)}_{vv\rho}}{\sinh2\rho} & \frac{f^{L(3)}_{v\rho\rho}}{\sinh^22\rho} \\
  \end{array} \right)_{\mu\nu},\\
 \label{3rdDL3}
 \Big(\bar{L}_{-1}L_{-1}\Big)^3 \Phi^{L,new}_{\rho\mu\nu}(u,v,\rho)
 &=&\frac{e^{-6t}}{\sinh^6\!\rho}e^{ikv}(\tanh\!\rho)^{ik}
  \left(
  \begin{array}{ccc}
     \frac{f^{L(3)}_{uu\rho}}{\sinh2\rho} & \frac{f^{L(3)}_{uv\rho}}{\sinh2\rho} & \frac{f^{L(3)}_{u\rho\rho}}{\sinh^22\rho}\\
    \frac{f^{L(3)}_{uv\rho}}{\sinh2\rho}  & \frac{f^{L(3)}_{vv\rho}}{\sinh2\rho} & \frac{f^{L(3)}_{v\rho\rho}}{\sinh^22\rho} \\
   \frac{f^{L(3)}_{u\rho\rho}}{\sinh^22\rho} &  \frac{f^{L(3)}_{v\rho\rho}}{\sinh^22\rho}  & \frac{f^{L(3)}_{\rho\rho\rho}}{\sinh^32\rho} \\
  \end{array}
 \right)_{\mu\nu}.
 \ea
Here again, the full expressions of $f^{L(3)}_{uuu},\cdots $ , are
written down in Appendix A2. The fourth descendants are given in
Appendix A3 with $s$-mode ($k=0$).

From these expressions, one can deduce the expression for  higher order of the descendants as
 \ba
  \Phi^{L(n),new}_{\rho\mu\nu}(u,v,\rho)&=&  \Big(\bar{L}_{-1}L_{-1}\Big)^n
                                      \Phi^{L,new}_{\rho\mu\nu}(u,v,\rho) \nonumber\\
 &=&\frac{e^{-2nt}}{\sinh^{2n}\!\rho}e^{ikv}(\tanh\!\rho)^{ik}F^{L(n)}_{\rho\mu\nu}(\rho),
 \label{lqnms}
 \ea
where $F^{L(n)}_{\rho\mu\nu}(\rho)$ is the corresponding $n$-th
order matrix.  As a result, we read off  the  left-logarithmic
quasinormal frequencies  of a traceless spin-3 field  from the
quasinormal modes (\ref{lqnms})
 \be
 \omega^n_L=-k-2in, ~~n\in N.
 \ee
 which is the same expression for spin-2 graviton
 $h_{\mu\nu}$~\cite{Sachs}. This is one of our main results.

It is by now appropriate to comment on the right-moving
solution and the right-logarithmic quasinormal modes. The
right-moving solution and its corresponding logarithmic solution
can be easily constructed by the substitution of both
$u\leftrightarrow v$, $L\leftrightarrow R$ and $\phi\rightarrow
-\phi$, $k\rightarrow -k$ in Eqs. (\ref{leftsol01}),
(\ref{leftm1}), and (\ref{nsol}). Moreover, the succeeding
descendants of the right-logarithmic quasinormal modes can also be
derived by the mentioned substitution, and finally yield the
quasinormal frequencies as
 \be
 \omega^n_R=k-2in, ~~n\in N.
 \ee


\subsection{Log-boundary conditions}
Since the time dependent part of the solution (\ref{lqnms}) is
simply given by exponential fall-off in $t$ as $ [e^{-2nt}]$ whereas
the radial part is a complicated form for each descendant, it would
be better to observe their asymptotic behaviors. For this purpose,
let us find the asymptotic behaviors of the left-logarithmic
solutions (\ref{nsol}). We have the asymptotic form in the
$\rho\rightarrow\infty$ limit as
 \ba
 \Phi^{L(0), new,\infty}_{u\mu\nu}(u,v,\rho)&\sim& \left(
  \begin{array}{ccc}
    0 & 0 & 0\\
    0 & 0 & 0 \\
    0 & 0 & 0 \\
  \end{array}
 \right)_{\mu\nu},  \nonumber\\
 \Phi^{L(0), new,\infty}_{v\mu\nu}(u,v,\rho)&\sim& -\rho\left(
  \begin{array}{ccc}
    0 & 0 & 0\\
    0 & 1 & e^{-2\rho} \\
    0 & e^{-2\rho} & e^{-4\rho} \\
  \end{array}
 \right)_{\mu\nu}, \nonumber\\
 \Phi^{L(0), new,\infty}_{\rho\mu\nu}(u,v,\rho)&\sim& -\rho\left(
  \begin{array}{ccc}
    0 & 0 & 0\\
    0 &  e^{-2\rho} & e^{-4\rho}\\
    0 & e^{-4\rho} & e^{-6\rho} \\
  \end{array}
 \right)_{\mu\nu}. \label{la0}
 \ea
The second component $\Phi^{L(0), new,\infty}_{v\mu\nu}$ in Eq.
(\ref{la0}) takes the same form as that of the spin-2
graviton~\cite{Sachs}. We point out that since the mode of
$\Phi^{L(0),new,\infty}_{vvv}(\propto \rho)$ is growing in $\rho$,
it could not be considered as a quasinormal mode. It may be cured by
taking descendants.  For example, taking the third descendants, we
have
 \ba
 \Phi^{L(3),new,\infty}_{u\mu\nu}(u,v,\rho)&\sim& -\rho\left(
  \begin{array}{ccc}
    e^{-6\rho} & e^{-4\rho} & e^{-6\rho}\\
    e^{-4\rho} & e^{-2\rho}  & e^{-4\rho} \\
    e^{-6\rho} & e^{-4\rho} & e^{-6\rho} \\
  \end{array}
 \right)_{\mu\nu},  \nonumber\\
 \Phi^{L(3),new,\infty}_{v\mu\nu}(u,v,\rho)&\sim& -\rho\left(
  \begin{array}{ccc}
    e^{-4\rho} & e^{-2\rho} & e^{-4\rho}\\
    e^{-2\rho} & -\frac{1}{\rho} & e^{-2\rho} \\
    e^{-4\rho} & e^{-2\rho} & e^{-4\rho} \\
  \end{array}
 \right)_{\mu\nu}, \nonumber\\
 \Phi^{L(3),new,\infty}_{\rho\mu\nu}(u,v,\rho)&\sim& -\rho\left(
  \begin{array}{ccc}
   e^{-6\rho} & e^{-4\rho} & e^{-6\rho}\\
    e^{-4\rho} & e^{-2\rho} & e^{-4\rho} \\
    e^{-6\rho} & e^{-4\rho} & e^{-6\rho} \\
  \end{array}
 \right)_{\mu\nu}. \label{thirdaym}
 \ea
We also would like to mention that all higher order components
$\{\Phi^{L(n),new,\infty}_{vvv} \sim 1\}$ with $n>0$ are not
dominant at large $\rho$, which implies that it may induce
difficulty in identifying the corresponding dual operator on  the
LCFT side.  On the other hand, all other components
$\{\Phi^{L(n),new,\infty}_{\rho\mu\nu}\}$ with $n\ge0$ for
$\rho,\mu,\nu\not=v$ show the exponential fall-off in $\rho$ as
$[\rho\cdot e^{-2 c \rho}]$ with $c=2,4,6$, which indicates
genuine gravitational quasinormal modes.

Finally,  it seems appropriate to comment on the fourth descendants
$\Phi^{L(4),new,\infty}_{\lambda\mu\nu}(u,v,\rho)$. Their asymptotic
behavior is exactly the same with the asymptotic form
(\ref{thirdaym}) of the third descendants. Thus, we expect that all
higher order descendants with $n>4$ for the spin-3 case behave as
like the third descendants have.

We have also proven that as were shown in Appendix B, these
properties persist to the noncritical cases of $\mu\not=\pm 1$.


\section{AdS/LCFT correspondence}
The log gravity at the chiral point could be dual to a LCFT on the
boundary described by $(\tau,\sigma)$~\cite{GJ1,Myunglcft}. In this
section, we show how to derive quasinormal frequencies
$\omega^n_{L/R}=\mp k-2 i n $ of the spin-3 field from the LCFT$_L$
on the boundary. It was known that the spin-3 chiral gravity with
the  Brown-Henneaux boundary condition~\cite{Brown:1986nw} is
holographically dual to the CFT$_L$  with classical $W_3$ algebra
and central charge $c_L=3l/G$~\cite{BLSS}. However, this is not our
case because we did not require the Brown-Henneaux boundary
condition.

The LCFT$_L$ ~\cite{Flohr,ML,Lewis}  may arise from  the two
operators $C$ and $D$  which satisfy the degenerate eigenequations
of $L_0$ as \be L_0 |C>=h_L|C>,~~~ L_0|D>=h_L|D>+|C>.
\label{lcdef} \ee The two-point functions of these operators take
the forms
\begin{eqnarray}\label{c11}
<C(x) C(0)>=0,~ <C(x) D(0)>=\frac{c}{x^{2h_L}},~ <D(x)
D(0)>=\frac{1}{x^{2h_L}}\left[d-2c\log(x)\right].
\end{eqnarray}
We note that Eq. (\ref{c11}) does not fix $C$ and its logarithmic
partner $D$  uniquely. For example, $D'=D+ a C$ also satisfies
Eq. (\ref{lcdef}). This freedom could be used to adjust the constant $d$
to any suitable value.

In order to derive quasinormal modes,  we focus at the location of
of the poles in the momentum space  for  the retarded two-point
functions $G^{CC}_R(\tau,\sigma)$, $G^{CD}_R(\tau,\sigma)$ and
$G^{DD}_R(\tau,\sigma)$~\cite{Sachs}. It is very important to
recognize that $G^{CD}_R(\tau,\sigma)$ is identical with that of the
two point function in the CFT~\cite{BSS}. The momentum space
representation can be read off from the commutator whose pole
structure is given by
\begin{eqnarray}
&&{\cal D}^{DC}(p_+) \propto \Gamma\left(h_L+i\frac{p_+}{2\pi
T_L}\right) \Gamma\left(h_L-i\frac{p_+}{2\pi T_L}\right), \label{9}
\end{eqnarray}
where $h_L=(m-2)/2$, $p_+=(\omega+ k)/2$, and
$T_L=r_+/2\pi=l\sqrt{M}/2\pi=1/2\pi$ for the nonrotating BTZ black
hole with $M=1$ and $l=1$. This function has poles in both the
upper and lower half of the $\omega$-plane. It turned out that the
poles located in the lower half-plane are the same as the poles of
the retarded two-point function $G^{CD}_R(\tau,\sigma)$.
Restricting the poles in Eq. (\ref{9}) to the lower half-plane, we
find one set of simple poles
\begin{eqnarray}
\omega_L&=&-k-2 i (n+h_L), \label{10}
\end{eqnarray}
with $n\in N$. This  set of poles characterizes the decay of the
perturbation on the LCFT$_L$ side.  Furthermore,
$G^{DD}_R(t,\sigma)$ can be inferred by noting~\cite{ML,Lewis} \be
<D(x) D(0)>=\frac{\partial}{\partial h_L}<C(x) D(0)>.
\ee
Then, this implies  that its momentum space representation takes the form
\begin{eqnarray}\label{ddpole}
{\cal D}^{DD}(p_+) \propto  \Gamma'\left(h_L+ip_+\right)
\Gamma\left(h_L-ip_+\right) +\Gamma\left(h_L+ip_+\right)
\Gamma'\left(h_L-ip_+\right),
\end{eqnarray}
where the prime ($'$) denotes the differentiation with respect to
$h_L$.  The poles in the lower half plane are relevant to deriving
quasinormal modes. We mention that  ${\cal D}^{DD}_R(p_+)$ has
double poles, while ${\cal D}^{CD}_R(p_+)$ has simple poles at the
same location.  These double poles  are responsible for the
linear-time dependence in $y(t,\rho)[=-t-\ln \sinh \rho]$ of the
corresponding quasinormal modes.

Now, we are in a position to  assign the bulk perturbation to  $C$ and
$D$.  For $m>2$,  the spin-3 perturbation $\Phi^L_{\rho\mu\nu}(m)$
is dual to a non-degenerate boundary operator $C$ with conformal
weight $h_L=\frac{m-2}{2}$.   At the chiral point of  $m=2$,  the
perturbation of $\Phi^L_{\rho\mu\nu}(m)|_{m\to 2}$ becomes a pure
gauge. In this case, Eq. (\ref{lcdef}) together with the relation of
$L_1D=0$  implies for the corresponding bulk perturbation
\cite{Lewis} \be
\Phi^{L,new}_{\rho\mu\nu}=[y(t,\rho)+a]\Phi^L_{\rho\mu\nu}(m)|_{m\to2},
\ee which shows that $\Phi^{L,new}_{\rho\mu\nu}$ is the bulk
perturbation for the logarithmic partner $D$. The bulk-boundary
correspondence is summarized as \be \Phi^L_{\rho\mu\nu}(m)|_{m\to 2}
\longleftrightarrow C,~~
\Phi^{L,new}_{\rho\mu\nu}\longleftrightarrow D. \ee

Similarly, we have the bulk-boundary correspondence for the
right-movers as \be \Phi^R_{\rho\mu\nu}(m)|_{m\to 2}
\longleftrightarrow \tilde{C},~~
\Phi^{R,new}_{\rho\mu\nu}\longleftrightarrow \tilde{D} \ee if one
introduces the right sector of LCFT$_R$ operator $\tilde{C}$ and
$\tilde{D}$ which satisfy the degenerate eigenequations for
$\bar{L}_0$ as \be \bar{L}_0 |\tilde{C}>=h_R|\tilde{C}>,~~~
\bar{L}_0|\tilde{D}>=h_R|\tilde{D}>+|\tilde{C}>. \label{2cdef} \ee
Here the two-point functions of these operators take the forms
\begin{eqnarray}\label{c22}
<\tilde{C}(\bar{x}) \tilde{C}(0)>=0,~ <\tilde{C}(\bar{x})
\tilde{D}(0)>=\frac{\bar{c}}{\bar{x}^{2h_R}},~ <\tilde{D}(\bar{x})
\tilde{D}(0)>=\frac{1}{\bar{x}^{2h_R}}\left[\bar{d}-2\bar{c}\log(\bar{x})\right].
\end{eqnarray}

On the other hand, its momentum space two-point functions take the form
\begin{eqnarray}
&&\bar{\cal D}^{\tilde{D}\tilde{C}}(p_-) \propto
\Gamma\left(h_R+i\frac{p_-}{2\pi T_R}\right)
\Gamma\left(h_R-i\frac{p_-}{2\pi T_R}\right) \label{12}
\end{eqnarray}
where $h_R=(m-2)/2$, $p_-=(\omega-k)/2$, and $T_R=1/2\pi$.
Confining the poles in Eq. (\ref{12}) to the lower half-plane, one
finds the other set of simple poles
\begin{eqnarray}
\omega_R&=&k-2 i (n+h_R). \label{13}
\end{eqnarray}
 This  set of poles characterizes the decay of the
perturbation on the LCFT$_R$ side.

Finally, we wish to mention that the above AdS/LCFT construction is
closely related to the spin-2 case since the bulk-boundary
correspondence is irrelevant to the higher spin $N$. Therefore, we
suggest that the spin-2 computations of 2- and 3-point
correlators~\cite{GS}, and the 1-loop partition
function~\cite{Gaberdiel:2010xv} may be helpful to calculate those
of spin-3.


\section{Discussions}
Using the operator method, we have constructed  the  logarithmic
quasinormal modes of a traceless spin-3 field  around the BTZ
black hole  at the critical point of the spin-3 TMG.  The
quasinormal frequencies  are given by $\omega^n_{L/R}=\mp k-2 in$.
The positive integer ``$n$" implies  that the BTZ black hole is
stable against the spin-3 perturbations because there is no
exponentially growing modes like $e^{2nt}$.  We note that these
quasinormal frequencies are the same as those of the spin-2
graviton.

The logarithmic quasinormal modes depending $y(t,\rho)=-(t+\ln \sinh
\rho)$ reflect that the linearized equation (\ref{spin3eq}) is a
third-order differential equation. The presence of the log spin-3
mode may induce  the instability of the BTZ black hole spacetimes
and the non-chiral nature of the spin-3 field coupled to
topologically massive gravity.  As far as the instability issue
concerned, we have to pay  attention  to  the log spin-2 case of
\cite{GJ1} where  the Brown-Henneaux boundary conditions are relaxed
to allow metric fluctuations to grow linearly in $\rho$ at infinity.
Also,  if one relaxes the boundary condition to allow log-modes
(whose presence makes the theory
non-chiral)~\cite{Grumiller:2008es,Henneaux:2009pw}, one has the
well-defined logarithmic quasinormal modes.     Moreover, even
though there is a linearized instability,  logarithmic excitations
always obey log-boundary conditions, but  not the Brown-Henneaux
boundary conditions~\cite{Maloney:2009ck}. At this stage,   it is
important to mention that the $y(t,\rho)$-dependence of the spin-3
quasinormal modes is necessary  to reproduce the double poles in
Eq.~(\ref{ddpole}).  In this direction,  we wish to note that the
appearance of a simple pole in the retarded Green function is
closely related to quasinormal modes~\cite{BSS}, while  the
appearance of a double pole in the retarded Green function  reflects
logarithmic quasinormal modes~\cite{Sachs}. Hence, the instability
due to $y(t,\rho)$ is not considered as an obstacle to the
interpretation of log spin-3 quasinormal modes.

Finally, we have established the bulk-boundary correspondence by
introducing two sets of operators ($D,C$) for the LCFT$_{L}$ and
($\tilde{D},\tilde{C}$) for the LCFT$_{R}$.   We could  read off the
logarithmic quasinormal frequencies of  $\omega^n_{L/R}$ from the
locations of retarded green function in momentum space ${\cal
D}^{DC}(p_+)$ and $\bar{\cal D
}^{\tilde{D}\tilde{C}}(p_-)$~\cite{BSS}.  These are
$\omega^{LCFT,n}_{L/R}=\mp k -2i (n+ h_{L/R})$ where
$h_{L/R}(m)=(m-2)/2$ is zero for $m=2$, leading to
$\omega^n_{L/R}=\omega^{LCFT,n}_{L/R}$. It dictates that using the
AdS/LCFT correspondence~\cite{Sachs}, the logarithmic quasinormal
frequencies could be obtained from the LCFT$_{L/R}$ sides.

\section*{Acknowledgement}
Two of us (Y. S. Myung and Y.-W. Kim) were supported by the
National Research Foundation of Korea (NRF) grant funded by the
Korea government (MEST) (Grant No.2011-0027293). Y.-J. Park was
partially supported by the National Research Foundation of Korea
(NRF) Grant funded by the Korea government (MEST) through the
Center for Quantum Spacetime (CQUeST) of Sogang University with
Grant No. 2005-0049409, and was also supported by World Class
University program funded by the Ministry of Education, Science
and Technology through the National Research Foundation of Korea
(Grant No. R31-20002).

\section*{Appendix: Full forms of the descendants}
\subsection*{A1. The second descendants of the left-logarithmic modes}

The full form of the components of the second descendants
(\ref{2ndDL1})-(\ref{2ndDL3}) are
 \ba
 f^{L(2)}_{uuv} &=& 14-4ik+2(12-7ik-k^2)y(t,\rho), \nonumber\\
 f^{L(2)}_{uu\rho} &=& 2f^{L(2)}_{uuv}= 2(14-4ik+2(12-7ik-k^2)y(t,\rho)), \nonumber\\
 f^{L(2)}_{uvv}&=& 40-29ik-5k^2+(26-11ik-k^2)\cosh2\rho \nonumber\\
               &+& 2\left(24-26ik-9k^2+ik^3+(12-7ik-k^2)\cosh2\rho\right)y(t,\rho), \nonumber\\
 f^{L(2)}_{uv\rho}&=& 2\left(40-29ik-5k^2+(40-15ik-k^2)\cosh2\rho\right.\nonumber\\
                   &+& \left.2(24-26ik-9k^2+ik^3+(24-14ik-2k^2)\cosh2\rho)y(t,\rho)\right),  \nonumber\\
 f^{L(2)}_{u\rho\rho}&=& 4\left(40-29ik-5k^2+(54-19ik-k^2)\cosh2\rho\right.\nonumber\\
                   &+& \left.2(24-26ik-9k^2+ik^3+(36-21ik-3k^2)\cosh2\rho)y(t,\rho)\right),  \nonumber\\
 f^{L(2)}_{vvv}&=& \frac{1}{8}\left(452-517ik-195k^2+24ik^3+4(88-81ik-23k^2+2ik^3)\cosh2\rho\right.\nonumber\\
               &+& \left.(12-7ik-k^2)\cosh4\rho\right)\nonumber\\
               &+& (48-76ik-44k^2+11ik^3+k^4+(24-26ik-9k^2+ik^3)\cosh2\rho)y(t,\rho), \nonumber\\
 f^{L(2)}_{vv\rho}&=& \frac{1}{4}\left(584-569ik-199k^2+24ik^3+4(168-139ik-33k^2+2ik^3)\cosh2\rho\right.\nonumber\\
               &+& \left.(88-43ik-5k^2)\cosh4\rho\right)\nonumber\\
               &+& (132-173ik-91k^2+22ik^3+2k^4+(144-156ik-54k^2+6ik^3)\cosh2\rho\nonumber\\
               &+& (12-7ik-k^2)\cosh4\rho)y(t,\rho), \nonumber\\
 f^{L(2)}_{v\rho\rho}&=& \frac{1}{2}\left(772-637ik-203k^2+24ik^3+4(248-197ik-43k^2+2ik^3)\cosh2\rho\right.\nonumber\\
               &+& \left.(220-95ik-9k^2)\cosh4\rho\right)\nonumber\\
               &+& 4(96-104ik-48k^2+11ik^3+k^4+(120-130ik-45k^2+5ik^3)\cosh2\rho\nonumber\\
               &+& (24-14ik-2k^3)\cosh4\rho)y(t,\rho), \nonumber\\
 f^{L(2)}_{\rho\rho\rho}&=& \left(1016-721ik-207k^2+24ik^3+ 4(328-255ik-53k^2+2ik^3)\cosh2\rho\right.\nonumber\\
              &+&       \left.(408-163ik-13k^2)\cosh4\rho\right)\nonumber\\
              &+& 4(276-257ik-103k^2+22ik^3+2k^4\nonumber\\
              &+& (336-364ik-126k^2+14ik^3)\cosh2\rho\nonumber\\
              &+& (108-63ik-9k^2)\cosh4\rho)y(t,\rho).
 \ea

\subsection*{A2. The third descendants of the left-logarithmic modes}

The full form of the components of the third descendants
(\ref{3rdDL1})-(\ref{3rdDL3}) are
 \ba
 f^{L(3)}_{uuu} &=& 6(47-24ik-3k^2+(60-47ik-12k^2+ik^3)y(t,\rho)), \nonumber\\
 f^{L(3)}_{uuv} &=& 3\left(342-285ik-78k^2+7ik^3+(248-143ik-24k^2+ik^3)\cosh2\rho\right.\nonumber\\
                &+&  \left. 2( 180-201ik-83k^2+15ik^3+k^4 +2(60-47ik-12k^2+ik^3)\cosh2\rho )y(t,\rho) \right), \nonumber\\
 f^{L(3)}_{uu\rho}&=& 6\left(342-285ik-78k^2+7ik^3+(342-191ik-30k^2+ik^3)\cosh2\rho\right.\nonumber\\
                &+&  \left. 2(180-201ik-83k^2+15ik^3+k^4 +3(60-48ik-12k^2+ik^3)\cosh2\rho )y(t,\rho) \right),  \nonumber\\
 f^{L(3)}_{uvv}&=& \frac{3}{8}\left(5714-6161ik-2522k^2+463ik^3+32k^4\right.\nonumber\\
                &+& 4(1386-1257ik-400k^2+51ik^3+2k^4)\cosh2\rho\nonumber\\
                &+& (582-379ik-78k^2+5ik^3)\cosh4\rho \nonumber\\
                &+&   \left(2(2460-3367ik-1860k^2+517ik^3+72k^4-4ik^5) \right.\nonumber\\
                &+&  24(180-201ik-83k^2+15ik^3+k^4)\cosh2\rho\nonumber\\
                &+& \left.\left.6(60-47ik-12k^2+ik^3)\cosh4\rho \right)y(t,\rho) \right), \nonumber\\
 f^{L(3)}_{uv\rho}&=& \frac{3}{4}\left(6894-6829ik-2630k^2+467ik^3+32k^4\right.\nonumber\\
                &+& 4(2070-1827ik-556k^2+65ik^3+2k^4)\cosh2\rho
                    + 9(154-95ik-18k^2+ik^3)\cosh4\rho \nonumber\\
                &+&   \left(2(3060-3837ik-1980k^2+527ik^3+72k^4-4ik^5) \right.\nonumber\\
                &+& 40(180-201ik-83k^2+15ik^3+k^4)\cosh2\rho\nonumber\\
                &+& \left.\left. 18(60-47ik-12k^2+ik^3)\cosh4\rho \right)y(t,\rho) \right),  \nonumber\\
 f^{L(3)}_{u\rho\rho}&=& \frac{3}{2}\left(8450-7689ik-2762k^2+471ik^3+32k^4\right.\nonumber\\
                &+& 4(2754-2397ik-712k^2+79ik^3+2k^4)\cosh2\rho\nonumber\\
                &+& (2566-1523ik-270k^2+13ik^3)\cosh4\rho \nonumber\\
                &+&   \left(2(3900-4495ik-2148k^2+541ik^3+72k^4-4ik^5) \right.\nonumber\\
                &+& 56(180-201ik-83k^2+15ik^3+k^4)\cosh2\rho\nonumber\\
                &+& \left.\left. 38(60-47ik-12k^2+ik^3)\cosh4\rho \right)y(t,\rho) \right), \nonumber\\
 f^{L(3)}_{vvv}&=& \frac{1}{16}\left(6(7014-9724ik-5417k^2+1516ik^3+213k^4-12ik^5)\right.\nonumber\\
                &+&  3(14268-17683ik-8476k^2+1941ik^3+208k^4-8ik^5)\cosh2\rho\nonumber\\
                &+& 6(882-888ik-327k^2+52ik^3+3k^4)\cosh4\rho + (60-47ik-12k^2+ik^3)\cosh6\rho\nonumber\\
                &+& \left(4(7380-12561ik-8947k^2+3411ik^3+733k^4-84ik^5-4k^6) \right.\nonumber\\
                &+&  48(540-783ik-450k^2+128ik^3+18k^4-ik^5)\cosh2\rho \nonumber\\
                &+& \left.\left.12(180-201ik-83k^2+15ik^3+k^4)\cosh4\rho \right)y(t,\rho) \right), \nonumber
 \ea
 \ba
 f^{L(3)}_{vv\rho}&=& \frac{1}{4}\left(3(10470-12808ik-6373k^2+1632ik^3+217k^4-12ik^5)\right.\nonumber\\
                &+&  3(13293-15287ik-6817k^2+1437ik^3+136k^4-4ik^5)\cosh2\rho\nonumber\\
                &+& 3(2970-2832ik-971k^2+140ik^3+7k^4)\cosh4\rho\nonumber\\
                &+& (441-307ik-69k^2+5ik^3)\cosh6\rho\nonumber\\
                &+& \left(2(11700-17385ik-10939k^2+3771ik^3+757k^4-84ik^5-4k^6) \right.\nonumber\\
                &+&  3(9540-13233ik-7380k^2+2063ik^3+288k^4-16ik^5)\cosh2\rho \nonumber\\
                &+& 30(180-201ik-83k^2+15ik^3+k^4)\cosh4\rho\nonumber\\
                &+& \left.\left.3(60-47ik-12k^2+ik^3)\cosh6\rho\right)y(t,\rho)\right),\nonumber\\
 f^{L(3)}_{v\rho\rho}&=& \frac{1}{4}\left(6(15294-17032ik-7641k^2+1776ik^3+221k^4-12ik^5)\right.\nonumber\\
                &+&  3(42256-45373ik-19104k^2+3819ik^3+336k^4-8ik^5)\cosh2\rho\nonumber\\
                &+& 6(6426-5916ik-1927k^2+256ik^3+11k^4)\cosh4\rho \nonumber\\
                &+& (3552-2321ik-480k^2+31ik^3)\cosh6\rho\nonumber\\
                &+& \left(4(18180-24621ik-13927k^2+4311ik^3+793k^4-84ik^5-4k^6) \right.\nonumber\\
                &+&  12(8220-10759ik-5748k^2+1565ik^3+216k^4-12ik^5)\cosh2\rho \nonumber\\
                &+& 156(180-201ik-83k^2+15ik^3+k^4)\cosh4\rho  \nonumber\\
                &+& \left.\left.36(60-47ik-12k^2+ik^3)\cosh6\rho\right)y(t,\rho)\right), \nonumber\\
 f^{L(3)}_{\rho\rho\rho}&=& \left(3(21486-22396ik-9221k^2+1948ik^3+225k^4-12ik^5)\right.\nonumber\\
                &+&  3(31203-31328ik-12479k^2+2388ik^3+200k^4-4ik^5)\cosh2\rho\nonumber\\
                &+& 15(2250-2028ik-639k^2+80ik^3+3k^4)\cosh4\rho \nonumber\\
                &+& (4599-2872ik-555k^2+32ik^3)\cosh6\rho\nonumber\\
                &+& \left(2(26820-34269ik-17911k^2+5031ik^3+841k^4-84ik^5-4k^6) \right.\nonumber\\
                &+&  3(25740-31683ik-16092k^2+4237ik^3+576k^4-32ik^5)\cosh2\rho \nonumber\\
                &+& 150(180-201ik-83k^2+15ik^3+k^4)\cosh4\rho  \nonumber\\
                &+& \left.\left.57(60-47ik-12k^2+ik^3)\cosh6\rho\right)y(t,\rho)\right).
 \ea

\subsection*{A3. The fourth descendent of the left-logarithmic modes}

Here we summarize the fourth descendent of the left-logarithmic
modes for the s-mode ($k=0$ case)
 \be
 \Phi^{L(4),new}_{\rho\mu\nu}(u,v,\rho)= \Big(\bar{L}_{-1}L_{-1}\Big)^4
 \Phi^{L,new}_{\rho\mu\nu}(u,v,\rho).
 \ee
Explicitly, we have
 \ba
 \Big(\bar{L}_{-1}L_{-1}\Big)^4 \Phi^{L,new}_{u\mu\nu}(u,v,\rho)
 &=&\frac{e^{-8t}}{\sinh^8\!\rho}
  \left(
  \begin{array}{ccc}
    f^{L(4)}_{uuu} & f^{L(4)}_{uuv} & \frac{f^{L(4)}_{uu\rho}}{\sinh2\rho}\\
    f^{L(4)}_{uuv} & f^{L(4)}_{uvv} & \frac{f^{L(4)}_{uv\rho}}{\sinh2\rho} \\
    \frac{f^{L(4)}_{uu\rho}}{\sinh2\rho} & \frac{f^{L(4)}_{uv\rho}}{\sinh2\rho} & \frac{f^{L(4)}_{u\rho\rho}}{\sinh^22\rho} \\
  \end{array}  \right)_{\mu\nu}, \\
 \Big(\bar{L}_{-1}L_{-1}\Big)^4 \Phi^{L,new}_{v\mu\nu}(u,v,\rho)
 &=&\frac{e^{-8t}}{\sinh^8\!\rho}
  \left(
  \begin{array}{ccc}
    f^{L(4)}_{uuv}  & f^{L(4)}_{uvv} & \frac{f^{L(4)}_{uv\rho}}{\sinh2\rho}\\
    f^{L(4)}_{uvv} & f^{L(4)}_{vvv} & \frac{f^{L(4)}_{vv\rho}}{\sinh2\rho} \\
    \frac{f^{L(4)}_{uv\rho}}{\sinh2\rho} & \frac{f^{L(4)}_{vv\rho}}{\sinh2\rho}  & \frac{f^{L(4)}_{v\rho\rho}}{\sinh^22\rho} \\
  \end{array} \right)_{\mu\nu},\\
 \Big(\bar{L}_{-1}L_{-1}\Big)^4 \Phi^{L,new}_{\rho\mu\nu}(u,v,\rho)
 &=&\frac{e^{-8t}}{\sinh^8\!\rho}
  \left(
  \begin{array}{ccc}
     \frac{f^{L(4)}_{uu\rho}}{\sinh2\rho} & \frac{f^{L(4)}_{uv\rho}}{\sinh2\rho} & \frac{f^{L(4)}_{u\rho\rho}}{\sinh^22\rho}\\
    \frac{f^{L(4)}_{uv\rho}}{\sinh2\rho}  & \frac{f^{L(4)}_{vv\rho}}{\sinh2\rho} & \frac{f^{L(4)}_{v\rho\rho}}{\sinh^22\rho} \\
   \frac{f^{L(4)}_{u\rho\rho}}{\sinh^22\rho} &  \frac{f^{L(4)}_{v\rho\rho}}{\sinh^22\rho}  & \frac{f^{L(4)}_{\rho\rho\rho}}{\sinh^32\rho} \\
  \end{array}
 \right)_{\mu\nu},
 \ea
where
 \ba
  f^{L(4)}_{uuu}&=&432(86+67\cosh2\rho)+8640(4+3\cosh2\rho)y(t,\rho),\nonumber\\
  f^{L(4)}_{uuv}&=&432(239+255\cosh2\rho+35\cosh4\rho)+2160(39+40\cosh2\rho+5\cosh4\rho)y(t,\rho),\nonumber\\
  f^{L(4)}_{uu\rho}&=&432(1+\cosh2\rho)(427+255\cosh2\rho+40(9+5\cosh2\rho)y(t,\rho)),\nonumber\\
  f^{L(4)}_{uvv}&=&36(5050+5955\cosh2\rho+1194\cosh4\rho+61\cosh6\rho)\nonumber\\
                &+& 1080(120+135\cosh2\rho+24\cosh4\rho+\cosh6\rho)y(t,\rho),\nonumber\\
  f^{L(4)}_{uv\rho}&=&(1+\cosh2\rho)\left(\right.144(2385+2068\cosh2\rho+199\cosh4\rho)\nonumber\\
                   &+& 17280(15+12\cosh2\rho+\cosh4\rho)y(t,\rho)\left.\right),\nonumber\\
  f^{L(4)}_{u\rho\rho}&=&2(1+\cosh2\rho)\left(\right.144(2925+3292\cosh2\rho+481\cosh4\rho)\nonumber\\
                   &+& 4320(75+82\cosh2\rho+11\cosh4\rho)y(t,\rho)\left.\right),\nonumber\\
  f^{L(4)}_{vvv}&=&\frac{9}{8}(186125+223320\cosh2\rho+47268\cosh4\rho+2792\cosh6\rho+15\cosh8\rho)\nonumber\\
                &+& 1080(120+135\cosh2\rho+24\cosh4\rho+\cosh6\rho)y(t,\rho),\nonumber\\
  f^{L(4)}_{vv\rho}&=&(1+\cosh2\rho)\left(\right.\frac{9}{2}(101650+97875\cosh2\rho+13086\cosh4\rho+349\cosh6\rho)\nonumber\\
                &+& 540(570+495\cosh2\rho+54\cosh4\rho+\cosh6\rho)y(t,\rho)\left.\right),\nonumber\\
  f^{L(4)}_{v\rho\rho}&=&2(1+\cosh2\rho)\left(\right.\frac{9}{2}(145670+177121\cosh2\rho+37066\cosh4\rho+1967\cosh6\rho)\nonumber\\
                &+& 4320(105+123\cosh2\rho+23\cosh4\rho+\cosh6\rho)y(t,\rho)\left.\right),\nonumber\\
  f^{L(4)}_{\rho\rho\rho}&=&2(1+\cosh2\rho)^2\left(\right.18(71495+65828\cosh2\rho+6717\cosh4\rho)\nonumber\\
                         &+& 6480(145+124\cosh2\rho+11\cosh4\rho)y(t,\rho)\left.\right).
 \ea


\subsection*{B. The asymptotic form of the left-moving noncritical modes}

The solution of the first order equation (\ref{meof}) is given by
 \be
 \Phi^L_{\rho\mu\nu}(u,v,\rho)=e^{-(\mu-1)(u+v)}(\sinh\rho)^{-2(\mu-1)}F^L_{\rho\mu\nu},
 \ee
where $\mu>1$. From this and the corresponding descendent
solutions, we find that the asymptotic forms are given as follows
 \ba
 \Phi^{L,\infty}_{u\mu\nu}(u,v,\rho)&\sim& e^{-2(\mu-1)t}\left(
  \begin{array}{ccc}
    0 & 0 & 0\\
    0 & 0 & 0 \\
    0 & 0 & 0 \\
  \end{array}
 \right)_{\mu\nu},  \nonumber\\
 \Phi^{L,\infty}_{v\mu\nu}(u,v,\rho)&\sim& e^{-2(\mu-1)t}\left(
  \begin{array}{ccc}
    0 & 0 & 0\\
    0 & e^{-2(\mu-1)\rho} & e^{-2\mu\rho} \\
    0 & e^{-2\mu\rho} & e^{-2(\mu+1)\rho} \\
  \end{array}
 \right)_{\mu\nu}, \nonumber\\
 \Phi^{L,\infty}_{\rho\mu\nu}(u,v,\rho)&\sim& e^{-2(\mu-1)t}\left(
  \begin{array}{ccc}
    0 & 0 & 0\\
    0 &  e^{-2\mu\rho} &  e^{-2(\mu+1)\rho}\\
    0 &  e^{-2(\mu+1)\rho} &  e^{-2(\mu+2)\rho} \\
  \end{array}
 \right)_{\mu\nu}.
 \ea
For the first descendent solutions, we have
 \ba
 \Phi^{L(1),\infty}_{u\mu\nu}(u,v,\rho)&\sim& e^{-2\mu t}\left(
  \begin{array}{ccc}
   0 & 0 & 0\\
    0 &  e^{-2\mu\rho} & e^{-2(\mu+1)\rho}\\
    0 & e^{-2(\mu+1)\rho} & e^{-2(\mu+2)\rho} \\
  \end{array}
 \right)_{\mu\nu},  \nonumber\\
 \Phi^{L(1),\infty}_{v\mu\nu}(u,v,\rho)&\sim& e^{-2\mu t}\left(
  \begin{array}{ccc}
    0 & e^{-2\mu\rho} & e^{-2(\mu+1)\rho}\\
    e^{-2\mu\rho} & e^{-2(\mu-1)\rho} & e^{-2\mu\rho} \\
   e^{-2(\mu+1)\rho} & e^{-2\mu\rho} & e^{-2(\mu+1)\rho} \\
  \end{array}
 \right)_{\mu\nu}, \nonumber\\
 \Phi^{L(1),\infty}_{\rho\mu\nu}(u,v,\rho)&\sim& e^{-2\mu t}\left(
  \begin{array}{ccc}
   0 & e^{-2(\mu+1)\rho} & e^{-2(\mu+2)\rho}\\
    e^{-2(\mu+1)\rho} & e^{-2\mu\rho} & e^{-2(\mu+1)\rho} \\
    e^{-2(\mu+2)\rho} & e^{-2(\mu+1)\rho} & e^{-2(\mu+2)\rho} \\
  \end{array}
 \right)_{\mu\nu}.
 \ea
For the second descendent solutions, we have
 \ba
 \Phi^{L(2),\infty}_{u\mu\nu}(u,v,\rho)&\sim& e^{-2(\mu+1)t}\left(
  \begin{array}{ccc}
    0 & e^{-2(\mu+1)\rho} & e^{-2(\mu+2)\rho}\\
    e^{-2(\mu+1)\rho} & e^{-2\mu\rho}  & e^{-2(\mu+1)\rho} \\
   e^{-2(\mu+2)\rho} & e^{-2(\mu+1)\rho} & e^{-2(\mu+2)\rho} \\
  \end{array}
 \right)_{\mu\nu},  \nonumber\\
 \Phi^{L(2),\infty}_{v\mu\nu}(u,v,\rho)&\sim& e^{-2(\mu+1)t}\left(
  \begin{array}{ccc}
    e^{-2(\mu+1)\rho}& e^{-2\mu\rho}  & e^{-2(\mu+1)\rho}\\
    e^{-2\mu\rho}  & e^{-2(\mu-1)\rho} & e^{-2\mu\rho}  \\
    e^{-2(\mu+1)\rho} & e^{-2\mu\rho}  & e^{-2(\mu+1)\rho} \\
  \end{array}
 \right)_{\mu\nu}, \nonumber\\
 \Phi^{L(2),\infty}_{\rho\mu\nu}(u,v,\rho)&\sim& e^{-2(\mu+1)t}\left(
  \begin{array}{ccc}
   e^{-2(\mu+2)\rho} &  e^{-2(\mu+1)\rho} & e^{-2(\mu+2)\rho}\\
    e^{-2(\mu+1)\rho} & e^{-2\mu\rho} &  e^{-2(\mu+1)\rho} \\
    e^{-2(\mu+2)\rho} &  e^{-2(\mu+1)\rho} & e^{-2(\mu+2)\rho} \\
  \end{array}
 \right)_{\mu\nu}.
 \ea
For the third descendent solutions, we have
 \ba\label{3rdncL}
 \Phi^{L(3),\infty}_{u\mu\nu}(u,v,\rho)&\sim& e^{-2(\mu+2)t}\left(
  \begin{array}{ccc}
    e^{-2(\mu+2)\rho} & e^{-2(\mu+1)\rho} & e^{-2(\mu+2)\rho}\\
    e^{-2(\mu+1)\rho} & e^{-2\mu\rho}  & e^{-2(\mu+1)\rho} \\
    e^{-2(\mu+2)\rho} & e^{-2(\mu+1)\rho} & e^{-2(\mu+2)\rho} \\
  \end{array}
 \right)_{\mu\nu},  \nonumber\\
 \Phi^{L(3),\infty}_{v\mu\nu}(u,v,\rho)&\sim& e^{-2(\mu+2)t}\left(
  \begin{array}{ccc}
     e^{-2(\mu+1)\rho} & e^{-2\mu\rho} &  e^{-2(\mu+1)\rho}\\
    e^{-2\mu\rho} & e^{-2(\mu-1)\rho}  & e^{-2\mu\rho} \\
     e^{-2(\mu+1)\rho} & e^{-2\mu\rho} &  e^{-2(\mu+1)\rho} \\
  \end{array}
 \right)_{\mu\nu}, \nonumber\\
 \Phi^{L(3),\infty}_{\rho\mu\nu}(u,v,\rho)&\sim& e^{-2(\mu+2)t}\left(
  \begin{array}{ccc}
  e^{-2(\mu+2)\rho} & e^{-2(\mu+1)\rho} & e^{-2(\mu+2)\rho}\\
    e^{-2(\mu+1)\rho}& e^{-2\mu\rho} & e^{-2(\mu+1)\rho} \\
    e^{-2(\mu+2)\rho} & e^{-2(\mu+1)\rho} & e^{-2(\mu+2)\rho} \\
  \end{array}
 \right)_{\mu\nu}.
 \ea
We note here that the fourth descendent of the solution has been
calculated for the s-mode ($k=0$ case), without loss of generosity.
As a result, we have confirmed that they are the same with the
asymptotic form of the third descendent Eq. (\ref{3rdncL}).


\end{document}